\documentclass[11pt,preprint]{aastex}

\usepackage{amsmath}

\shorttitle{High Ecliptic Latitude Survey}

\shortauthors{Terai, Takahashi, \& Itoh}

\begin{document}

\title{High Ecliptic Latitude Survey for Small Main-Belt Asteroids\altaffilmark{*}}

\altaffiltext{*}{Based on data collected at Subaru Telescope, which is operated by the National
Astronomical Observatory of Japan (NAOJ).
}

\author{Tsuyoshi Terai$^{1}$, Jun Takahashi$^{2}$, and Yoichi Itoh$^{2}$}
\affil{
$^1$National Astronomical Observatory of Japan, 2-21-1 Osawa, Mitaka, Tokyo 181-8588, Japan\\
$^2$Center for Astronomy, University of Hyogo, 407-2 Nishigaichi, Sayo-cho, Sayo-gun, Hyogo
679-5313, Japan\\
}
\email{tsuyoshi.terai@nao.ac.jp}

\begin{abstract}
Main-belt asteroids have been continuously colliding with one another since they were formed.
Its size distribution is primarily determined by the size dependence of asteroid strength against
catastrophic impacts.
The strength scaling law as a function of body size could depend on collision velocity, but the
relationship remains unknown especially under hypervelocity collisions comparable to
10~km~sec$^{-1}$.
We present a wide-field imaging survey at ecliptic latitude of around 25$\arcdeg$ for investigating
the size distribution of small main-belt asteroids which have highly inclined orbits.
The analysis technique allowing for efficient asteroid detections and high-accuracy photometric
measurements provide sufficient sample data to estimate the size distribution of sub-km asteroids
with inclinations larger than 14$\arcdeg$.
The best-fit power-law slopes of the cumulative size distribution is 1.25~$\pm$~0.03 in the
diameter range of 0.6--1.0~km and 1.84~$\pm$~0.27 in 1.0--3.0~km.
We provide a simple size distribution model that takes into consideration the oscillations of the
power-law slope due to the transition from the gravity-scaled regime to the strength-scaled regime.
We find that the high-inclination population has a shallow slope of the primary components of the size
distribution compared to the low-inclination populations.
The asteroid population exposed to hypervelocity impacts undergoes collisional processes that
large bodies have a higher disruptive strength and longer life-span relative to tiny bodies than
the ecliptic asteroids.

\end{abstract}


%%%%%%%%%%%%%%%%%%%%%%%%%%%%%%%%%%%%%%%%%%%%%%%%%%%%%%%%%%%%%%%%%%%%%%%%%%%%%%%%%%%%%%%%%%%%%%%%%%%
%%%%%%%%%%%%%%%%%%%%%%%%%%%%%%%%%%%%%%%%%%%%%%%%%%%%%%%%%%%%%%%%%%%%%%%%%%%%%%%%%%%%%%%%%%%%%%%%%%%
\section{INTRODUCTION}\label{sec:introduction}

Main-belt asteroids (MBAs) have continuously undergone self-collisional processes.
The impact events are characterized by the target/impactor masses and collision velocity.
When kinetic energy of an impactor is larger than the critical specific energy $Q^\ast_D$,
the energy per unit target mass required to shatter the target and disperse half of its mass,
the target is catastrophically disrupted \citep{Davis+02}.
Otherwise, the target (largest fragment) retains a mass larger than half of the original,
resulting cratering or gravitational reaccumulation of collisional fragments after shattering.
$Q^\ast_D$ is an indicator of impact strength, which depends on the body size
\citep[e.g.][]{Housen&Holsapple90,Durda+98,Benz&Asphaug+99}.
$Q^\ast_D$ decreases with increasing diameter for asteroids less than $\sim$0.1--1~km, called
`strength-scaled regime'.
In contrast, it increases with increasing diameter for the larger asteroids, called
`gravity-scaled regime'.
The degree of change in $Q^\ast_D$ with body size is the primary determinant of a power-law size
distribution of the small body population in a collisional cascade \citep{O'Brien&Greenberg03}.

\begin{figure}
\epsscale{.80}
\plotone{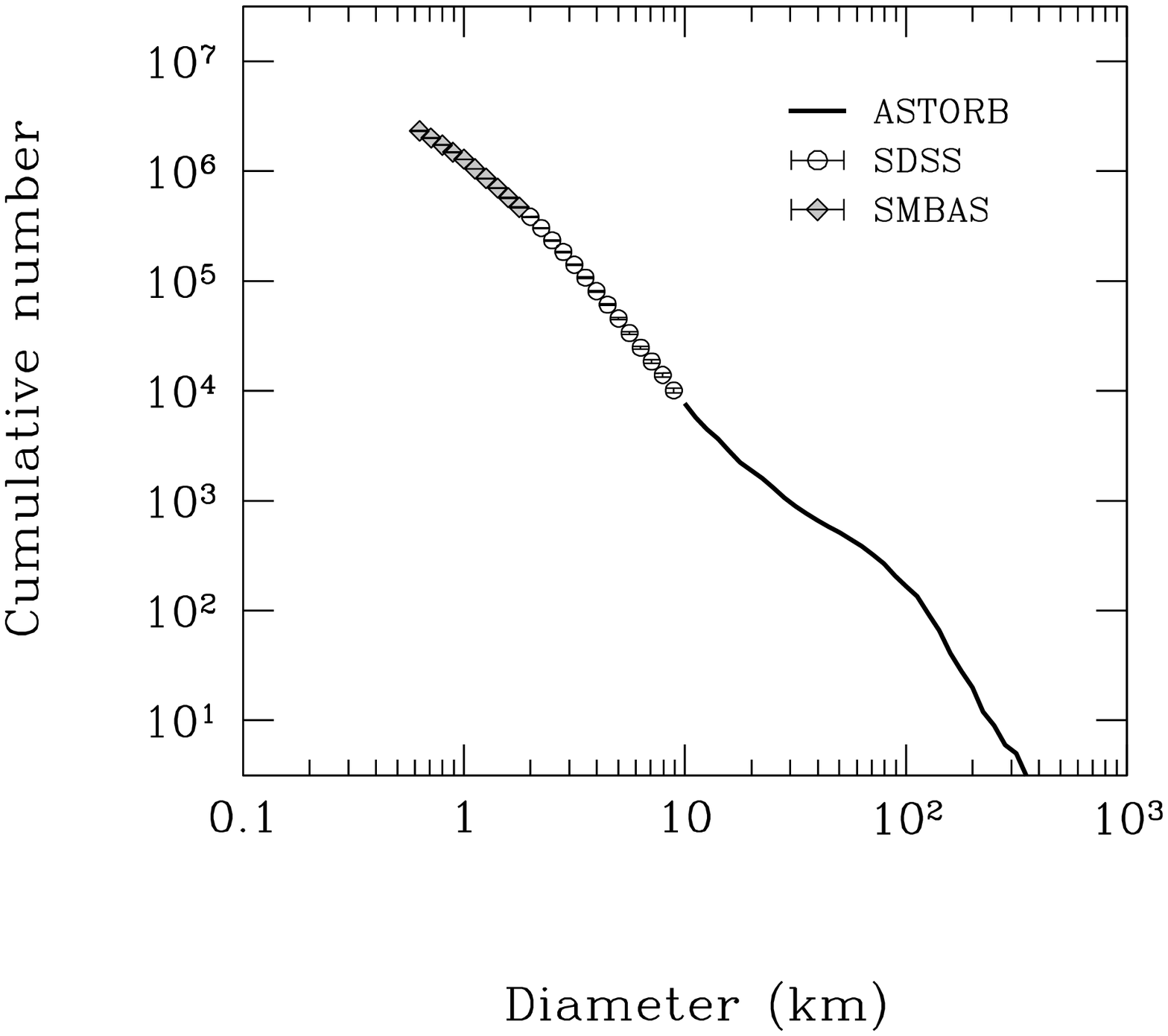}
\caption{
 The cumulative size distribution of main-belt asteroids estimated from the previous observations. 
 The solid line represents the population of observed asteroids listed in the ASTORB database
 \citep{Bowell+94}.
 The circles and diamonds represent extrapolations based on the results of the Sloan Digital Sky
 Survey \citep{Ivezic+01} and Subaru Main Belt Asteroid Survey \citep{Yoshida&Nakamura07},
 respectively.
 \label{fig1}}
\end{figure}

The size distribution of MBAs down to sub-km size has been estimated by previous extensive surveys
with ground-based telescopes such as Palomar-Leiden Survey \citep[PLS;][]{vanHouten+70},
Spacewatch \citep{Jedicke&Metcalfe+98}, Sloan Digital Sky Survey \citep[SDSS;][]{Ivezic+01},
Subaru Main Belt Asteroid Survey \citep[SMBAS;][]{Yoshida+03,Yoshida&Nakamura07}, and
Sub-Kilometer Asteroid Diameter Survey \citep[SKADS;][]{Gladman+09}.
In addition, infrared satellites including IRAS \citep{Tedesco+02}, AKARI \citep{Usui+11}, and
WISE \citep{Masiero+11} measured accurate diameters of numerous asteroids.
Figure~\ref{fig1} shows the cumulative size distribution of MBAs compiled from the Asteroid
Orbital Elements Database\footnote{$\url{\rm ftp://ftp.lowell.edu/pub/elgb/astorb.html}$}
\citep[ASTORB;][]{Bowell+94} and the survey results reported by SDSS and SMBAS.
Using the observed size distributions, many studies devoted efforts for modeling the collisional
evolution of MBAs via numerical simulations
\citep{Durda+98,Bottke+05a,Bottke+05b,O'Brien&Greenberg05,deElia&Brunini07}.
It should be noted that the $Q^\ast_D$ law is supposed to be a function only of asteroid size in
each model.

\cite{Petit+01} presented a dynamical evolution model for primordial asteroids in the early main
belt that were dynamically excited due to gravitational perturbations from Jupiter and embedded
planetary embryos.
In this phase, collisions between asteroids occurred at higher velocities than at present
\citep[$\sim$4~km~sec$^{-1}$;][]{Vedder+98} because of the pumped-up eccentricities and
inclinations \citep{Bottke+05b}.
\cite{Bottke+05a} pointed out that $Q^\ast_D$ law could be affected by varying collision
velocities.
They suggested a steeper $Q^\ast_D$ curve in the gravity-scaled regime for 10~km~sec$^{-1}$
collisions than that for slower collisions.

The asteroid collisional evolution among impacts with much higher velocities (i.e.
$\gg$~4~km~sec$^{-1}$; hereinafter called `hypervelocity') than the mean collisions in the main
belt remains unknown.
Because of the technical difficulties, only a few laboratory experiments for hypervelocity
collisions have been conducted \citep{Kadono+10,Takasawa+11}.
The hydrocode simulations by \cite{Benz&Asphaug+99} indicated that in the gravity-scaled regime,
$Q^\ast_D$ for a basalt target has similar slopes between collisions of 3~km~sec$^{-1}$ and
5~km~sec$^{-1}$, while $Q^\ast_D$ for a icy target in 3~km~sec$^{-1}$ collisions increases with
size more steeply than that in 0.5~km~sec$^{-1}$ collisions.
However, another study with impact simulations showed that the slope of $Q^\ast_D$ for a basalt
target in 5~km~sec$^{-1}$ collisions is shallower than that in 3~km~sec$^{-1}$ collisions
\citep{Jutzi+10}.
The collision-velocity dependency of the $Q^\ast_D$ law has not yet been confirmed.

Investigation of an asteroid population with highly inclined orbits is an effective means to
understand the material properties against hypervelocity collisions.
In the main belt, asteroids with orbits inclined at higher than 15$\arcdeg$ (hereinafter called
high-inclination MBAs) have the mean collision velocities exceeding $\sim$7~km~sec$^{-1}$
\citep{Farinella&Davis+92,Gil-Hutton06}.
Those high-inclination MBAs remain in the collisional evolution dominated by hypervelocity impacts.
The size distribution of high-inclination MBAs enables to examine the $Q^\ast_D$ law under
collisional processes at high velocity.

\cite{Terai&Itoh11} performed a survey focused on high-inclination small-size MBAs in 9.0-deg$^2$
fields using the data with a detection limit of $r$~=~24.0~mag obtained by 8.2-m Subaru Telescope.
They detected 178 MBA candidates with 0.7--7~km in diameter and found that the size distribution
of high-inclination MBAs is shallower than that of low-inclination MBAs over a wide diameter range
from 0.7~km to 50~km.
However, the faint-end slopes in diameter less than 2~km based on the own survey data
potentially include large bias due to the non-uniform data taken at sky regions with various
ecliptic latitudes and solar phase angles in uneven atmospheric conditions.
Actually, the power-law index of the size distribution for low-inclination MBAs is inconsistent
with that presented by previous studies \citep{Ivezic+01,Yoshida&Nakamura07}.

In this paper, we present the results of an additional survey to determine the size distribution of
high-inclination MBAs down to sub-km diameter.
We note that the size distribution of MBAs is poorly represented by a single power law.
As seen in figure~\ref{fig1}, the distribution has significant slope transitions around 3~km,
20~km, and 100~km in diameter, called `wavy' structure
\citep{CampoBagatin+94,Durda+98,Davis+02,O'Brien&Greenberg03}.
For accurate comparison of the power-law slope between the wavy-patterned size distributions,
measurements of the distribution shape in the diameter range from sub-km to several km is required
to determine the intrinsic slope.

We carried out an uniform wide-field imaging at high ecliptic latitudes using the Subaru Telescope.
This survey allows to obtain a large amount of homogeneous data which give three times more sample
of small high-inclination MBAs than the previous study.
We evaluate the difference of size distribution between low- and high-inclination MBAs considering
the wavy structure and taxonomic distribution.
The results provide useful clues for understanding the collisional evolution of primordial
asteroids in the early solar system, and also that of planetesimals in some debris disks that
have been found outside the solar system.

%%%%%%%%%%%%%%%%%%%%%%%%%%%%%%%%%%%%%%%%%%%%%%%%%%%%%%%%%%%%%%%%%%%%%%%%%%%%%%%%%%%%%%%%%%%%%%%%%%%
%%%%%%%%%%%%%%%%%%%%%%%%%%%%%%%%%%%%%%%%%%%%%%%%%%%%%%%%%%%%%%%%%%%%%%%%%%%%%%%%%%%%%%%%%%%%%%%%%%%
\section{ASTEROID SURVEY}\label{sec:survey}

%%%%%%%%%%%%%%%%%%%%%%%%%%%%%%%%%%%%%%%%%%%%%%%%%%%%%%%%%%%%%%%%%%%%%%%%%%%%%%%%%%%%%%%%%%%%%%%%%%%
\subsection{Observations}\label{sec:observations}

Our survey was performed on August 24 and 25, 2008 (UT) using the Suprime-Cam mounted on
the 8.2-m Subaru Telescope.
The Suprime-Cam is a mosaic camera with ten 2k~$\times$~4k CCD chips and covers a
34$\arcmin$~$\times$~27$\arcmin$ field of view with a pixel scale of 0.20$\arcsec$
\citep{Miyazaki+02}.
The data were taken at sky area centered on RA~(J2000)~=~21$^{\rm h}$40$^{\rm m}$ and
Dec~(J2000)~=~+14$\arcdeg$00$\arcmin$, within 6$\arcdeg$ from opposition in ecliptic longitude.
The region with ecliptic latitude of around 25$\arcdeg$ is suitable to detect asteroids with
inclination of 15$\arcdeg$ or higher in the main belt.
We imaged 104 fields which contain no bright background objects.
Most of the fields overlap the sky coverage of
SDSS Data Release 9\footnote{\url{http://www.sdss3.org/dr9/}} (DR9).
Each field was visited twice with 240-sec exposures at an interval of 20~min using the
$r$-band filter.
The seeing size is ranging 0$\farcs$7--1$\farcs$0 in almost all the data.
The total surveyed area is 26.5~deg$^2$.

%%%%%%%%%%%%%%%%%%%%%%%%%%%%%%%%%%%%%%%%%%%%%%%%%%%%%%%%%%%%%%%%%%%%%%%%%%%%%%%%%%%%%%%%%%%%%%%%%%%
\subsection{Data Analysis}\label{sec:analysis}

\begin{figure}
\epsscale{.50}
\plotone{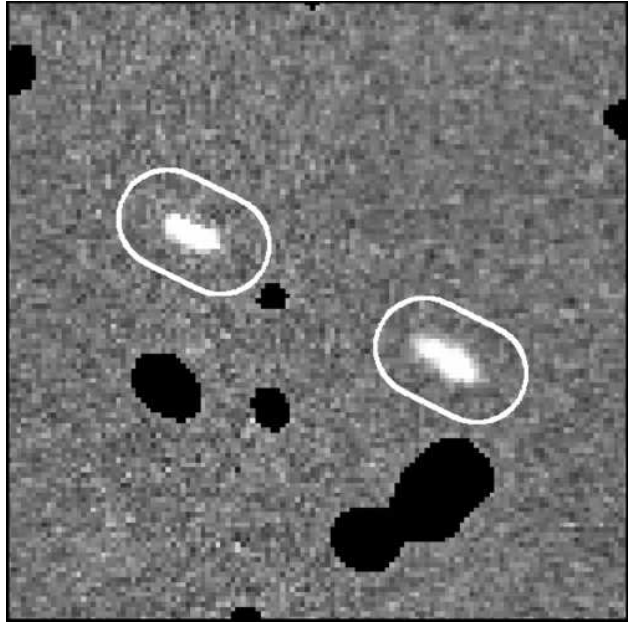}
\caption{
 Detected moving objects in the combined image from two-visit images.
 The original images were taken with 240-sec exposures at an interval of 20~min under a seeing size
 of 0.8~arcsec.
 The field of view covers 25$\arcsec$~$\times$~25$\arcsec$ with north up and east to the left.
 The background objects have been masked (black regions).
 The moving objects have a sky motion of 38~arcsec~hr$^{-1}$ and brightness of
 $r$~=~23.07~$\pm$~0.07~mag.
 The circles surrounding the objects show the apertures for photometry.
 \label{fig2}}
\end{figure}

The images were processed using IRAF produced by the National Optical Astronomy Observatories
(NOAO) and SDFRED2 \citep{Ouchi+04}.
The standard procedure of data reduction includes overscan subtraction, flat-fielding, correction
of geometric distortion, subtraction of sky background, and position matching between two images
which were taken at a same field.
Moving objects are searched by the image processing technique presented in \cite{Terai+07}.
The two-visit imaging with a 20-min interval allows to identify MBAs that have sky motion faster
than $\sim$30~arcsec~hr$^{-1}$ at near opposition (see figure~\ref{fig2}).

We measured the positions and brightness of detected moving objects using the SExtractor
\citep{Bertin&Arnouts96} and IRAF/APPHOT package, respectively.
In the images acquired with 240-sec exposures, most asteroids are trailed
as seen in figure~\ref{fig2}.
For precise photometry, we produced synthetic apertures appropriate to each object through the
following procedures.
(i) The circular aperture for absolute photometry of point sources in the image is determined.
Its radius is set to $\sim$2.5~times the full width at half-maximum (FWHM) of point sources.
(i\hspace{-.1em}i) The motion velocity of the moving object is estimated from the difference
between its central coordinates in the two-visit images.
(i\hspace{-.1em}i\hspace{-.1em}i) The circular aperture around the object's center is evenly
extended the distance that the object moved during the exposure time in the both directions along
the axis of the motion.
The `moving-circular' apertures formed in these ways are drawn in figure~\ref{fig2}.
We estimated the total flux within the given apertures using the POLYPHOT task.

We also conducted photometry on the field stars listed in the SDSS DR9 catalog with
$r$~=~19.0--20.0~mag in the AB system using the same technique for the flux calibration..
The magnitude zero-point of each data was determined from the measured total flux and $r$-band
magnitude provided in the catalog.
However, the filter transmission of the Suprime-Cam differs from that of SDSS.
The difference in $r$-band magnitude of the field stars, $r_{\rm sup}-r_{\rm sdss}$, was corrected
with the SDSS color ($r-i$)$_{\rm sdss}$ using the transformation equation presented by Fumiaki
Nakata,
\begin{equation}
r_{\rm sup}-r_{\rm sdss}=-0.00282-0.0498(r-i)_{\rm sdss}
-0.0149(r-i)_{\rm sdss}^2.
\label{eq_rmag}
\end{equation}
The ($r-i$)$_{\rm sdss}$ colors of the field stars were derived from the SDSS DR9 catalog.
The resulting zero point of the Suprime-Cam filter system was applied to calculate the apparent
magnitudes of detected moving objects in the frame.

\begin{figure}
\epsscale{.80}
\plotone{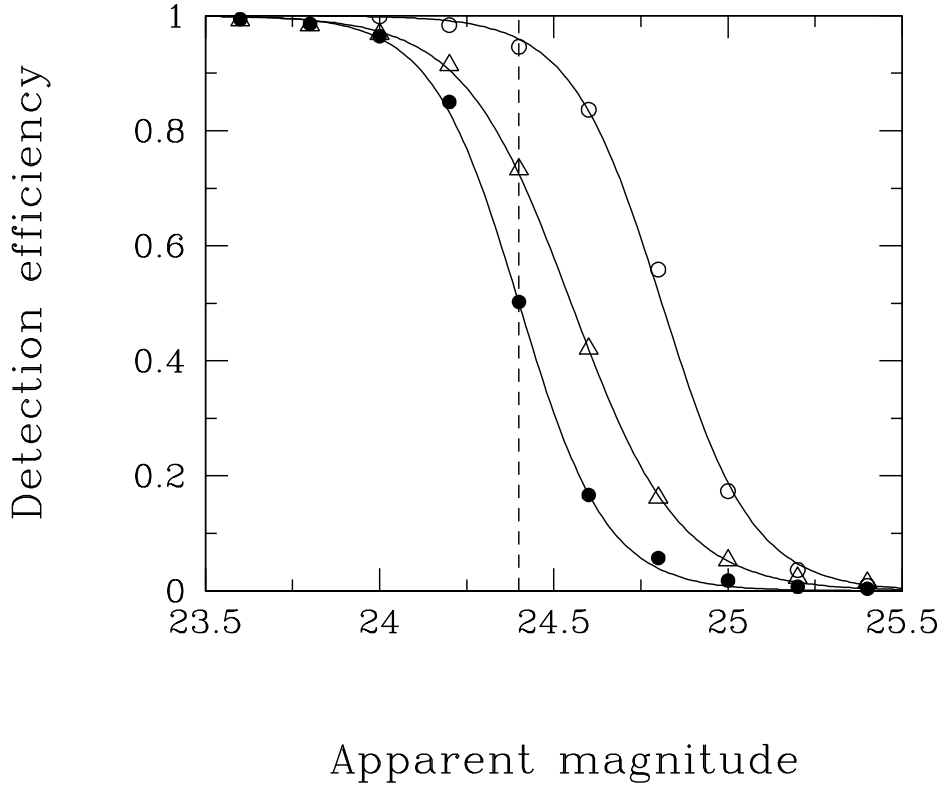}
\caption{
 Net detection efficiency $\eta^\prime(r)$ for moving objects as a function of apparent
 $r$ magnitude.
 The detection efficiency reduces with motion velocity due to the decrease in flux density.
 In this work, the motion velocity was set to 40~arcsec~hr$^{-1}$, similar to or faster than
 those of most of the detected MBA candidates (see figure~\ref{fig5} and \ref{fig6}).
 Only the data with $\eta^\prime(r)$~$\geq$~0.5 at $r$~=~24.4~mag was used for our analysis of
 the asteroid population.
 The filled circles and open circles represent objects with the minimum and maximum
 $\eta^\prime(r)$ at $r$~=~24.4~mag (dashed line) in the selected data, respectively.
 The open triangles show combined $\eta^\prime(r)$ of all the selected data.
 This curve indicates that objects even with close to $r$~=~24.4~mag can be detected with efficiencies of
 about 0.7 or higher in most of the images.
 The faintest objects in our sample cause little increase in statistical uncertainty of
 measurement of the size distributions.
 The solid curves are best-fit functions given by equation (\ref{eq_deteff}).
 \label{fig3}}
\end{figure}

\begin{figure}
\epsscale{.90}
\plotone{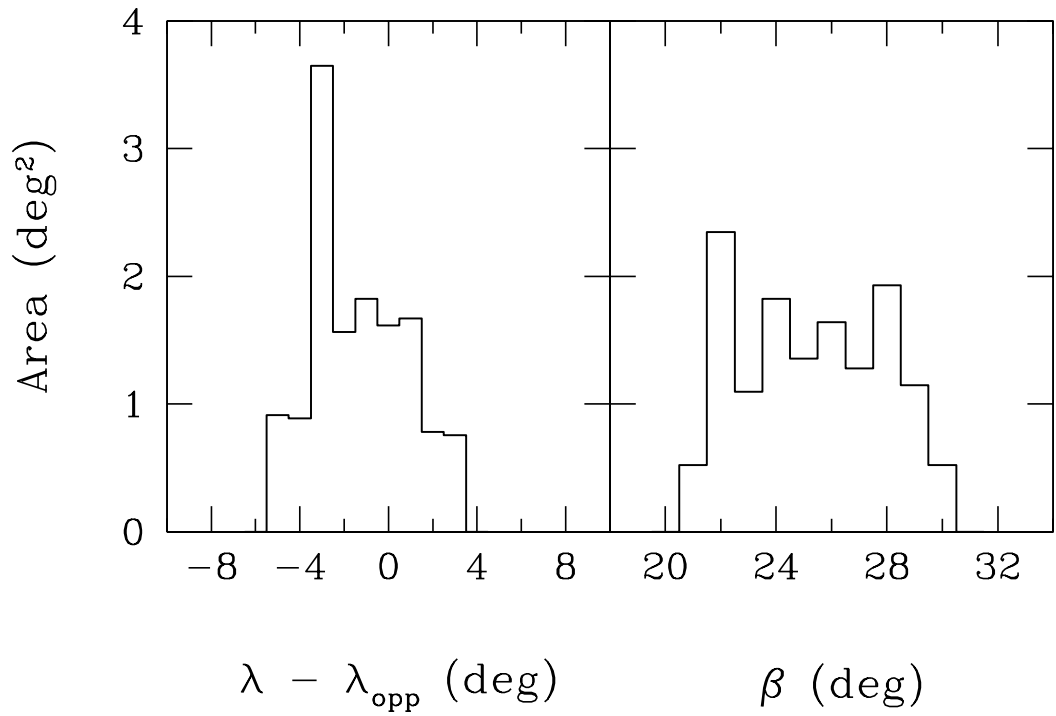}
\caption{
 Distributions of the ecliptic longitude with respect to opposition
 ($\lambda$~$-$~$\lambda_{\rm opp}$ ; left) and ecliptic latitude ($\beta$ ; right)
 covered by the selected data.
 \label{fig4}}
\end{figure}

To obtain the statistically homogeneous sample, we evaluated limiting magnitude of the all data.
The detection efficiency was estimated using artificial asteroid trails implanted into the raw data
with every 0.2~mag.
The fractions of detection were represented by
\begin{equation}
\eta(r) = \frac{A}{2} \left[ 1 - \tanh \left( \frac{r-m_{50}}{w} \right) \right],
\label{eq_deteff}
\end{equation}
where $A$, $m_{50}$, and $w$ are the maximum efficiency, half-maximum magnitude, and transition
width, respectively \citep{Gladman+98}.
Mean $A$ is 0.84 due to the sky coverage of background objects.
We defined $r$~=~24.4~mag as the limiting magnitude in this survey.
We excluded the data with the net detection efficiency $\eta^\prime(r)$ = $\eta(r)$/$A$ less
than 50\% at $r$~=~24.4~mag.
Figure~\ref{fig3} shows the $\eta^\prime(r)$ curves with the minimum (filled circles) and
maximum (open circles) values of the selected data at 24.4~mag.
The combined $\eta^\prime(r)$ of all the selected data are also plotted (open triangles).
The selected data cover 13.6~deg$^2$ in actual or 11.4~deg$^2$ in effect.
Figure~\ref{fig4} shows histograms of the ecliptic longitude from the opposition and ecliptic
latitude covered by the selected data.

%%%%%%%%%%%%%%%%%%%%%%%%%%%%%%%%%%%%%%%%%%%%%%%%%%%%%%%%%%%%%%%%%%%%%%%%%%%%%%%%%%%%%%%%%%%%%%%%%%%
%%%%%%%%%%%%%%%%%%%%%%%%%%%%%%%%%%%%%%%%%%%%%%%%%%%%%%%%%%%%%%%%%%%%%%%%%%%%%%%%%%%%%%%%%%%%%%%%%%%
\section{RESULTS}\label{sec:results}

\begin{figure}
\epsscale{.80}
\plotone{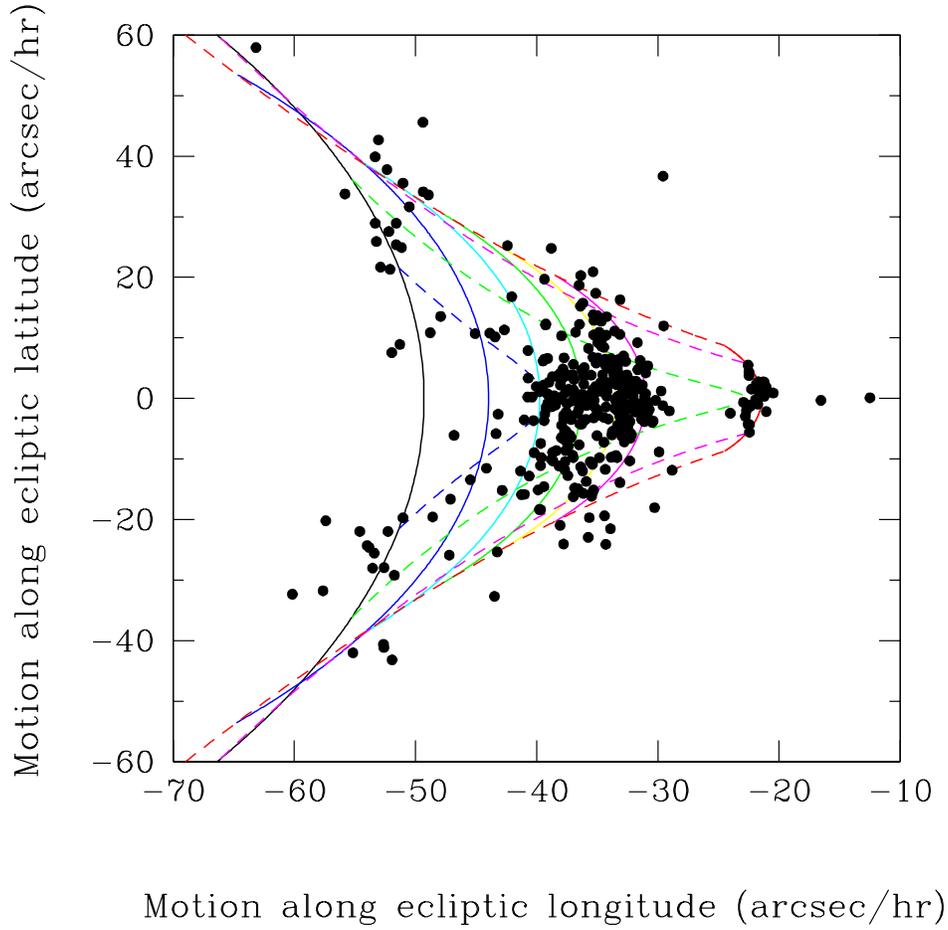}
\caption{
 Sky motion distribution in the ecliptic coordinate system for moving objects detected in this
 survey.
 The solid lines represent motions of asteroids in circular orbits with semi-major axes of
 1.8~AU (black), 2.1~AU (blue), 2.4~AU (cyan), 2.7~AU (green), 3.0~AU (yellow), 3.3~AU (magenta),
 and 5.2~AU (red) at the ecliptic longitude of opposition and latitude of +25$\arcdeg$.
 The dashed lines represent motions of asteroids in circular orbits with inclinations of
 15$\arcdeg$ (blue), 20$\arcdeg$ (green), 30$\arcdeg$ (magenta), and 40$\arcdeg$ (red) at the
 ecliptic longitude of opposition and latitude of +25$\arcdeg$.
 \label{fig5}}
\end{figure}

Our exploration found 441 moving objects in the selected data with the 50\% detection limit of
$r$~$>$~24.4~mag.
Figure~\ref{fig5} shows the distribution of their sky motions in the geocentric ecliptic
coordinate system.
The major group with the negative longitudinal motions of $\sim$30--45~arcsec~hr$^{-1}$ consists of
MBAs, while the clump around 22~arcsec~hr$^{-1}$ corresponds to Jovian Trojans.

At the region with the geocentric ecliptic latitude $\beta$ and longitude with respect to
opposition $\lambda^{\prime}$, the lower inclination limit of detectable asteroids, $I_{\rm lim}$,
is
\begin{equation}
 \sin I_{\rm lim} = \frac{\Delta}{R} \sin \phi,
 \label{eq_ilim}
\end{equation}
where $R$ is the heliocentric distance, $\Delta$ is the geocentric distance, and $\phi$ is the
elongation angle between the Sun and the asteroid given by
$\cos \phi = \cos \lambda^\prime \cos \beta$.
In the survey area of $\beta$ $\sim$ 25$\arcdeg$ near the opposition, $I_{\rm lim}$ is around
15$\arcdeg$.
Asteroids with inclination $I_{\rm lim}$ show no motion along the ecliptic latitude at the
opposition field.
In contrast, latitudinally-moving asteroids have inclinations higher than $I_{\rm lim}$.
The dotted lines in figure~\ref{fig5} represent the motions of asteroids in circular orbits
when they are observed at opposition ($\lambda^{\prime}$~=~0$\arcdeg$) and
$\beta$ = 25$\arcdeg$.

%%%%%%%%%%%%%%%%%%%%%%%%%%%%%%%%%%%%%%%%%%%%%%%%%%%%%%%%%%%%%%%%%%%%%%%%%%%%%%%%%%%%%%%%%%%%%%%%%%%
\subsection{Estimation of Orbital Parameters}\label{sec:orbit}

Two-visit positioning of moving objects in a night is insufficient for orbit determination.
Instead, we estimated semi-major axis and inclination of each asteroid from the sky motion assuming
that the orbit is circular, namely, the eccentricity is zero.
The adequacy of this assumption is evaluated below.
\cite{Jedicke96} presents the expressions of ecliptic motion derived from orbital elements and sky
coordinate \citep[see also the appendix in][]{Ivezic+01}.

Let us consider an asteroid in a circular orbit with semi-major axis $a$ and inclination $I$
located at a heliocentric ecliptic longitude with respect to opposition $l^{\prime}$ and latitude
$b$.
We use the coordinate system defined in figure 2 of \cite{Jedicke96}.
The position vector from the Sun, $\mathbf{R}$, and angular momentum vector, $\mathbf{h}$, are
given by
\begin{eqnarray}
&&\mathbf{R} = a \ (\cos l^{\prime} \cos b, \ \sin l^{\prime} \cos b, \ \sin b), \\
&&\mathbf{h} = \sqrt{\mu a} \ (\sin \Omega \sin I, \ -\cos \Omega \sin I, \ \cos I),
\end{eqnarray}
respectively, where $\mu$ is the product of the gravitational constant and mass of the Sun, and
$\Omega$ is the longitude of the ascending node derived from $l^{\prime}$, $b$, and $I$.
The relative velocity with respect to the Earth is given by
\begin{equation}
\mathbf{v} = -\frac{\mathbf{R} \times \mathbf{h}}{a^2} - (0, \sqrt{\mu}, 0).
\end{equation}
The observed ecliptic motion is converted from $\mathbf{v}$ using the unit vectors representing the
geocentric directions of increasing ecliptic longitude, latitude, and distance.
These vectors are given by
\begin{mathletters}
\begin{eqnarray}
&&\hat{\lambda}^{\prime}=\ (-\sin\lambda^{\prime}, \ \cos\lambda^{\prime}, \ 0), \\
&&\hat{\beta}=\ (-\cos\lambda^{\prime}\sin\beta, \ -\sin\lambda^{\prime} \sin\beta, \ \cos\beta).
\end{eqnarray}
\end{mathletters}
The rates of apparent motion are
\begin{mathletters}
\begin{eqnarray}
&& \dot{\lambda}^{\prime} = \frac{\mathbf{v}}{\Delta} \cdot \hat{\lambda}^{\prime},\\
&& \dot{\beta} = \frac{\mathbf{v}}{\Delta} \cdot \hat{\beta}.
 \label{eq_motion}
\end{eqnarray}
\end{mathletters}

\begin{figure}
\epsscale{.80}
\plotone{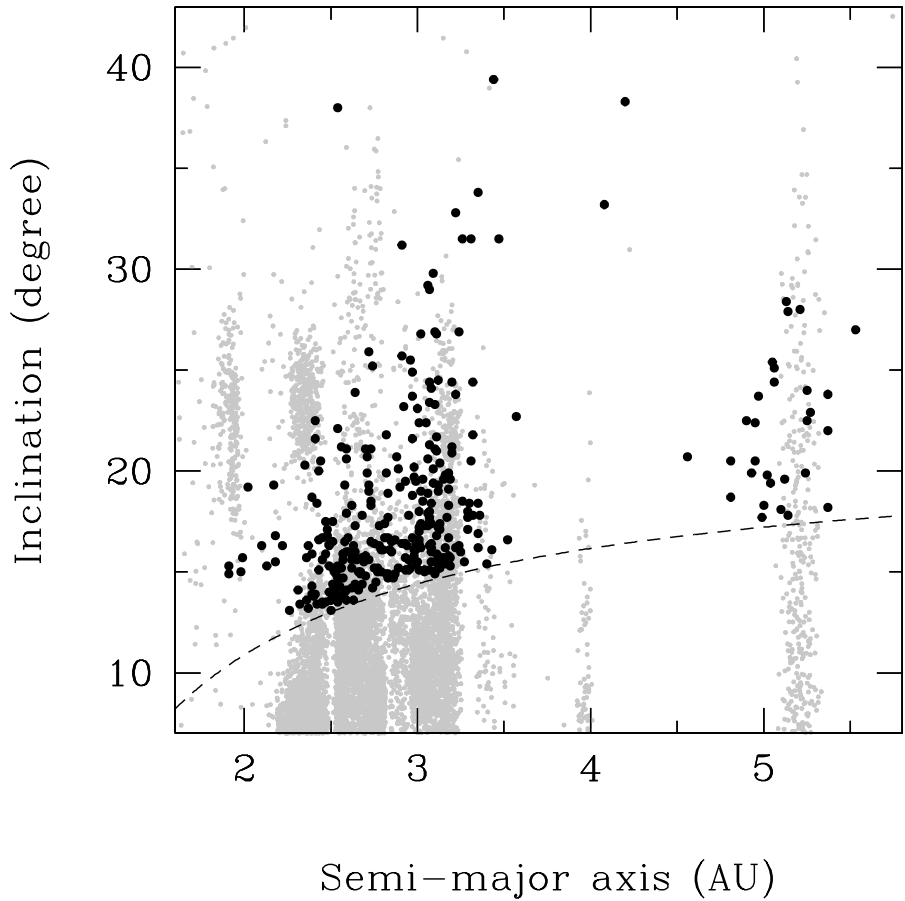}
\caption{
 Estimated semi-major axes and inclinations of moving objects detected in this
 survey assuming circular orbits (black dots).
 The major swarm consists of main-belt asteroids.
 The minor swarm around 5~AU corresponds to the Jovian Trojan group.
 The orbits of 50,000 known asteroids are also plotted as gray dots.
 The dashed curve shows the detection limit given by the lowest ecliptic latitude (+21$\arcdeg$)
 in the survey fields.
 \label{fig6}}
\end{figure}

The orbital elements of a detected moving object are derived from the best-fit set of $a$ and $I$
for equation~(\ref{eq_motion}).
Several objects with inclinations larger than 40$\arcdeg$ were excluded because of the significant
uncertainty of estimated $a$ and $I$ (see figure~\ref{fig5}).
Figure~\ref{fig6} shows the distribution of detected asteroids in the semi-major axis vs.
inclination space.
The dashed curve represents the inclination limit of detectable asteroids given by
equation~(\ref{eq_ilim}).
MBAs and Jovian Trojans can be identified clearly.
We put objects with $a$~=~2.0--3.3~AU into MBA candidates.

The estimation accuracy of orbital elements of the MBA candidates was evaluated by Monte-Carlo
simulation of a virtual asteroid survey.
10,000 hypothetical asteroids were randomly generated in the area of
-5$\arcdeg$~$<$~$\lambda^{\prime}$~$<$~+5$\arcdeg$ and +20$\arcdeg$~$<$~$\beta$~$<$~+30$\arcdeg$
(see figure~\ref{fig4}).
The orbit elements were given in the orbital parameter space ranging $a$~=~2.3--3.3~AU,
$e$~=~0.0--0.5, and $I$~=~15$\arcdeg$--40$\arcdeg$ in inclination.
These were governed by the probability distributions based on the orbital distribution of
known MBAs with $I$ $>$ 15$\arcdeg$ listed in the ASTORB database.
The simulation showed that the rms errors are 0.14~AU in semi-major axis and 4.5$\arcdeg$
in inclination, which are comparable to the values presented by \cite{Nakamura&Yoshida02} for an
ecliptic survey.
The estimated heliocentric distances have uncertainty of $\sim$0.37~AU.
We note that the systematic errors in heliocentric distance and inclination are less than
0.05~AU and 0.1$\arcdeg$, respectively, which are small enough to be negligible compared to the
random errors.

%%%%%%%%%%%%%%%%%%%%%%%%%%%%%%%%%%%%%%%%%%%%%%%%%%%%%%%%%%%%%%%%%%%%%%%%%%%%%%%%%%%%%%%%%%%%%%%%%%%
\subsection{Estimation of Asteroid Size}\label{sec:size}

The apparent $r$-band magnitude of the detected moving objects is converted into the absolute
magnitude, $H_r$, by
\begin{equation}
H_r = r - 5 \log(R\cdot\Delta) - P(\alpha),
 \label{eq_hmag}
\end{equation}
where $\alpha$ is the phase angle, the angle between the Sun and the Earth from the asteroid,
and $P(\alpha)$ is the phase function.
We used the $P(\alpha)$ expressions presented by \cite{Bowell+89} assuming the slope parameter
$G$~=~0.15.
The phase angles of MBA candidates range from 6$\arcdeg$ to 14$\arcdeg$.

Asteroid diameter $D$ in km is estimated from
\begin{equation}
\log D =  0.2r_\sun - \log \frac{\sqrt{p}}{2(\rm AU/km)} - 0.2H_r,
 \label{eq_diam}
\end{equation}
where $r_\sun$ and $p$ represent the $r$-band magnitude of the Sun in the AB system and the
geometric albedo.
In the SDSS photometric system, $r_\sun$~=~-26.91~mag and $(r - i)_\sun$~=~0.13~mag
\citep{Fukugita+11}, which are converted into the Suprime-Cam $r_\sun$ by equation~(\ref{eq_rmag}).

The geometric albedo was assigned the mean value of albedo-known asteroids.
MBAs mainly consist of two major groups: redder/brighter asteroids dominated by S-type
asteroids and bluer/darker asteroids dominated by C-type asteroids.
These are hereinafter called S-like asteroids and C-like asteroids, respectively.
We used the mean albedos obtained from the AKARI All-Sky Survey observations, 0.22 for the S-like
asteroids and 0.07 for the C-like asteroids \citep{Usui+11}.
The mean albedo of total asteroids depends on the number ratio between S- and C-type asteroids
which varies with heliocentric distance.

Assuming that the heliocentric distribution of each group is constant with asteroid size,
we estimated fractions of the S-/C-type asteroids from the SDSS Moving Object
Catalog\footnote{\url{http://www.astro.washington.edu/users/ivezic/sdssmoc/sdssmoc.html}}
\citep[SDSS MOC;][]{Ivezic+02}.
Ivezi\'c et al. (2001) divided asteroids into the two color groups using a color index given by
\begin{equation}
a^{\ast} = 0.89(g - r) + 0.45(r - i) - 0.57.
 \label{eq_aast}
\end{equation}
We classified the SDSS MOC asteroids with $a^{\ast}$~$>$~0 into S-like asteroids, and those with
$a^{\ast}$~$<$~0 into C-like asteroids.
The number ratios of the C-like to S-like derived from the orbit-known asteroids with
$I$~$>$~15$\arcdeg$ and $D$~$>$~5~km (larger than the limiting size of complete detection for the
both types) are 0.5 in the inner belt ranging $R$~=~2.0--2.5~AU, 1.7 in the middle belt ranging
$R$~=~2.5--3.0~AU, and 10.0 in the outer belt ranging $R$~=~3.0--3.3~AU.
The weighted mean albedos in the inner, middle, and outer belts are 0.17, 0.13, and 0.09,
respectively.

As seen in equation~(\ref{eq_hmag}), $H_r$ is derived from measurements of the apparent $r$-band
magnitude, $R$, and $\Delta$.
The apparent magnitude includes 1$\sigma$ uncertainties of $\sim$0.15~mag at the faint end, namely
$r \sim 24$~mag, where $\sigma$ is the standard deviation.
The estimated $R$ and $\Delta$ have uncertainties of $\sim$0.37~AU (see Section \ref{sec:orbit}).
These errors cause the uncertainty in $H_r$ of $\sim$0.7~mag.
It corresponds to a $\sim$30~\% error in $D$.

%%%%%%%%%%%%%%%%%%%%%%%%%%%%%%%%%%%%%%%%%%%%%%%%%%%%%%%%%%%%%%%%%%%%%%%%%%%%%%%%%%%%%%%%%%%%%%%%%%%
%%%%%%%%%%%%%%%%%%%%%%%%%%%%%%%%%%%%%%%%%%%%%%%%%%%%%%%%%%%%%%%%%%%%%%%%%%%%%%%%%%%%%%%%%%%%%%%%%%%
\section{DISCUSSION}\label{sec:discussion}

%%%%%%%%%%%%%%%%%%%%%%%%%%%%%%%%%%%%%%%%%%%%%%%%%%%%%%%%%%%%%%%%%%%%%%%%%%%%%%%%%%%%%%%%%%%%%%%%%%%
\subsection{Sample Selection}\label{sec:selection}

\begin{figure}
\epsscale{.80}
\plotone{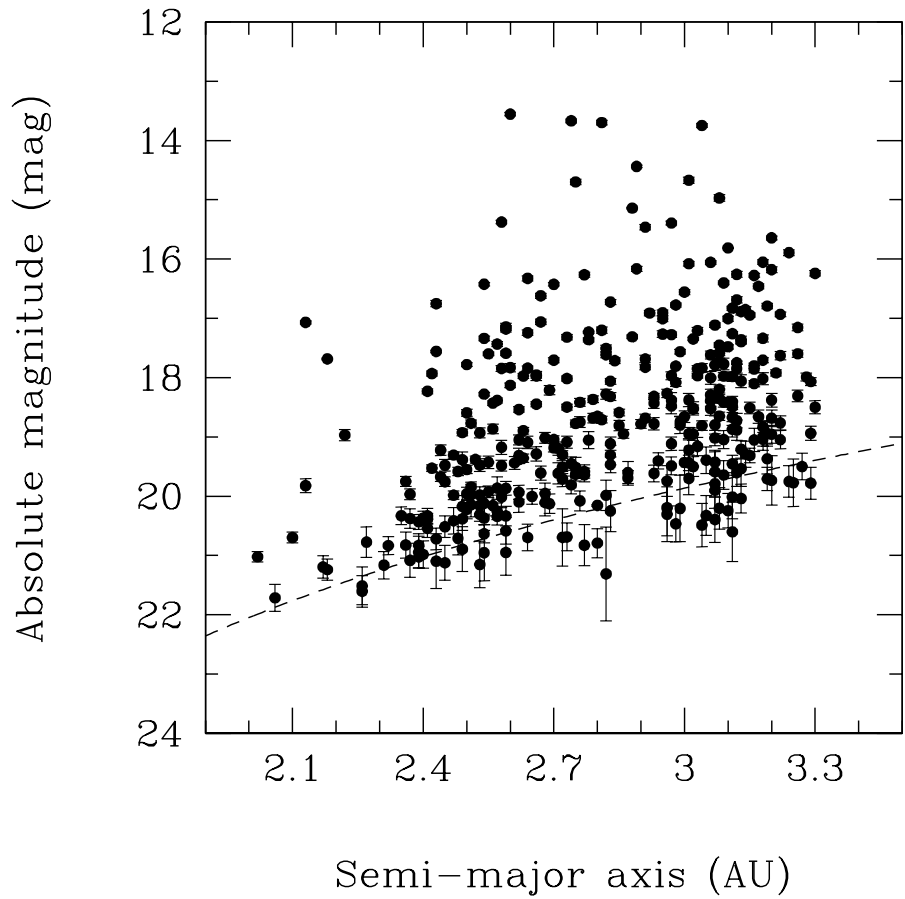}
\caption{
 Plot of estimated semi-major axis vs. absolute $r$ magnitude in the AB system for the main-belt
 asteroid candidates.
 The error bars represent photometric uncertainty.
 The dashed curve shows the 50\%-complete detection limit corresponding to apparent magnitude of
 $r$~=~24.4~mag.
 \label{fig7}}
\end{figure}

Figure \ref{fig7} shows the distribution in semi-major axis versus absolute magnitude of the MBA
candidates selected in section \ref{sec:orbit}.
The error bars display the 1$\sigma$ uncertainty in photometric measurements.
The dashed curve represents the limiting magnitude of 50\%-complete detection, namely
$r$~=~24.4~mag.
At the outer edge of the main belt defined as $R$~=~3.3~AU, the detection limit corresponds to
$H_r$~=~19.4~mag.
We put the MBA candidates with $H_r$~$\leq$~19.4~mag into the final sample including 221 objects
for statistical analysis.

\begin{figure}
\epsscale{1.00}
\plottwo{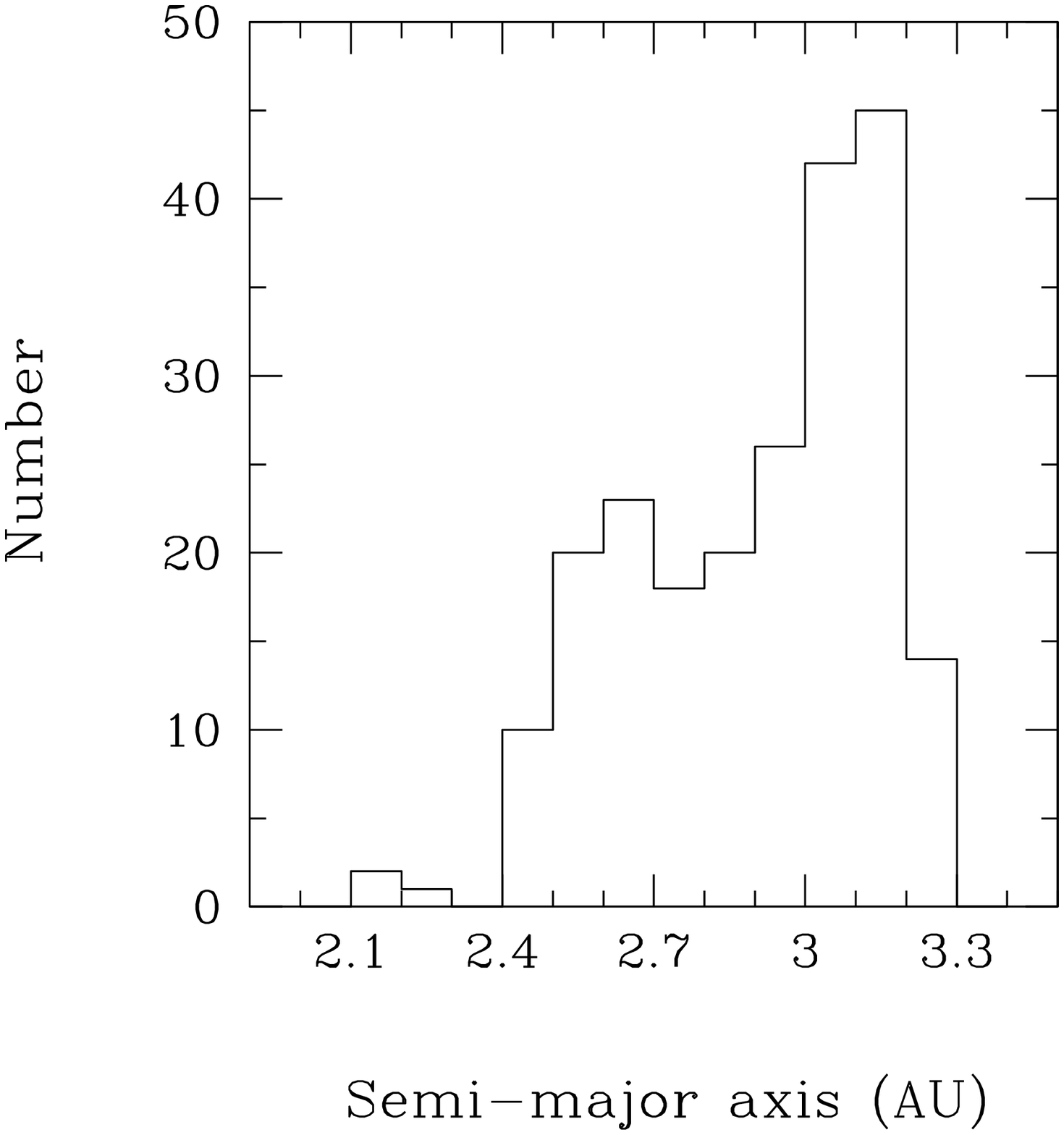}{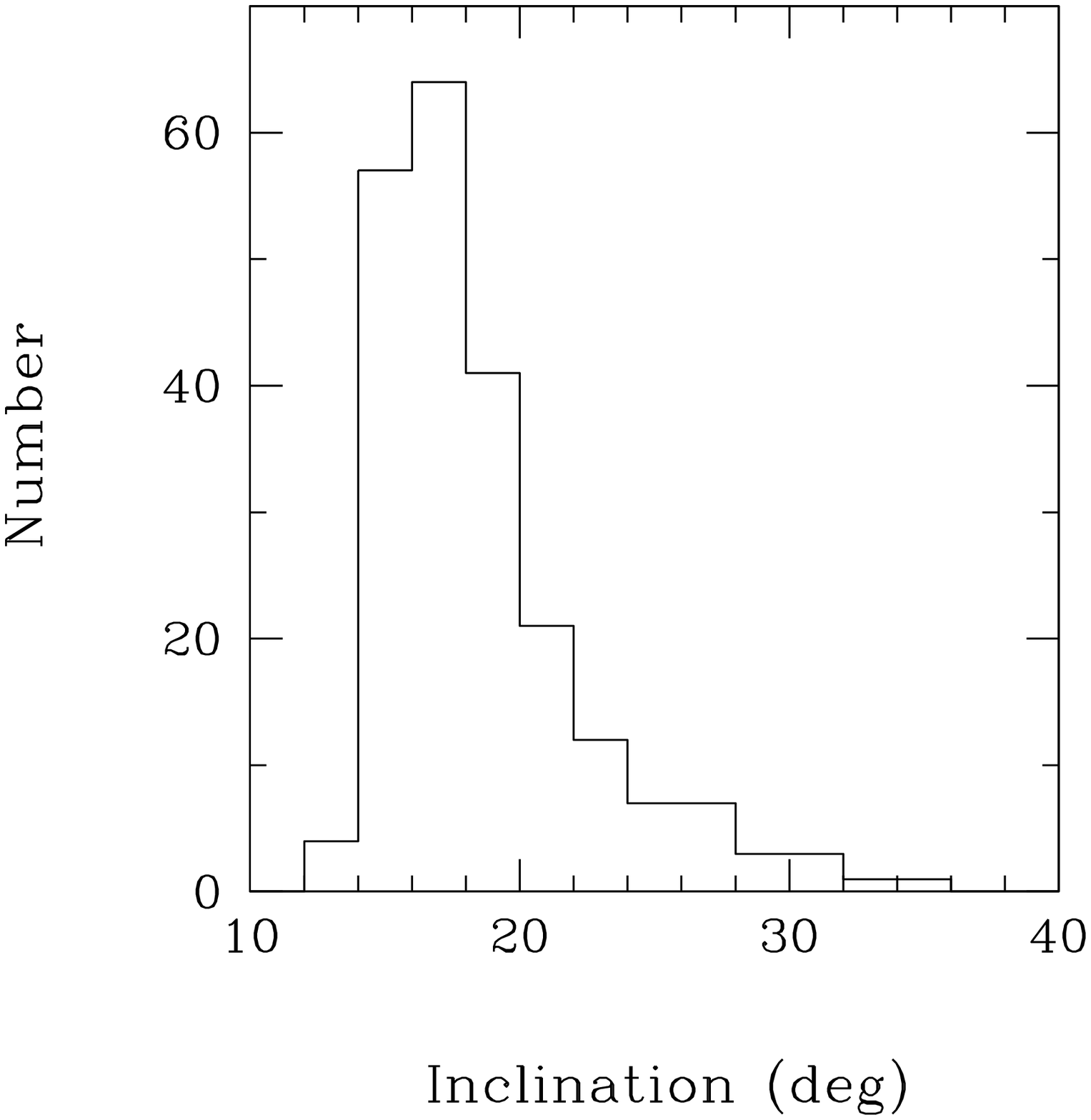}
\caption{
 Number distributions of estimated semi-major axis (left) and inclination (right) for the final
 sample of MBA candidates with $H_r$~$\leq$~19.4~mag ($H_r$ is absolute $r$ magnitude in the
 AB system).
 \label{fig8}}
\end{figure}

Figure \ref{fig8} shows the number distributions of semi-major axis and inclination for the
sample asteroids.
Almost all of them have inclinations larger than 14$\arcdeg$, allowing to examine the size
distribution of high-inclination MBAs.
Also, the sample asteroids are mostly located in the outer region of main belt and therefore are
likely to be dominated by C-type asteroids \citep{Bus&Binzel02}.

%%%%%%%%%%%%%%%%%%%%%%%%%%%%%%%%%%%%%%%%%%%%%%%%%%%%%%%%%%%%%%%%%%%%%%%%%%%%%%%%%%%%%%%%%%%%%%%%%%%
\subsection{Size Distribution}\label{sec:size_distribution}

\begin{figure}
\epsscale{.85}
\plotone{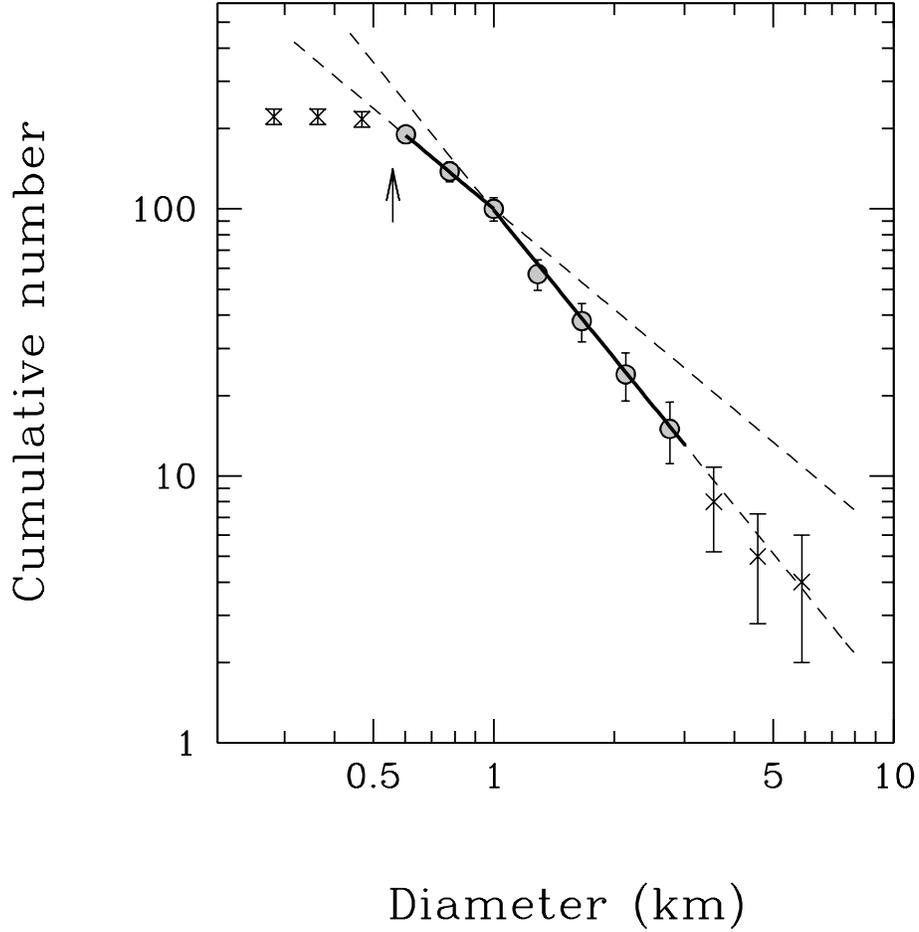}
\caption{
 Cumulative size distribution of the MBA sample.
 The cumulative number $N$($>$$D$), where $D$ is diameter in km, has been corrected for detection
 efficiency.
 The error bars represent $\sqrt N$, where $N$ is the cumulative number.
 The data points in the fitting regions are plotted as circles, while those excluded from the
 fitting are plotted as cross signs.
 The straight lines show the best-fit power-law functions in
 0.6~km $<$ $D$ $<$ 1.0~km and 1.0~km $<$ $D$ $<$ 3.0~km.
 The arrow shows the detection limit, $D$~=~0.56~km (see text).
 \label{fig9}}
\end{figure}

Figure \ref{fig9} shows the cumulative size distribution (CSD) as a function of asteroid
diameter obtained from the final sample.
The cumulative number was weighted by the net detection efficiency of each object as
$1/\eta^\prime(r)$  (see Section \ref{sec:analysis}).
The arrow represents the asteroid size converted from the 50\% detection limit at $R$~=~3.3~AU
($H_r$~=~19.4~mag) with the albedo given for the outer-belt asteroids ($p$~=~0.09), corresponding to
$D$~=~0.56~km.

The CSD is represented by a power-law expression as $\Sigma(>D) \propto D^{-b}$, where $\Sigma(>D)$
is the number of asteroids larger than $D$ in diameter per square degree.
The power-law index $b$ gives the CSD slope.
We found that the CSD has a knee around $D$~=~1.0~km, namely, the slope in $D$~$<$~1.0~km is
shallower than that in $D$~$>$~1.0~km.
This feature has also been indicated in the previous survey for low-inclination sub-km MBAs by
\cite{Yoshida&Nakamura07}.
Hence, we fixed two diameter regions, 0.6~km to 1.0~km and 1.0~km to 3.0~km, for characterization
of the CSD.
The region with $D$~$>$~3~km was excluded because of the large uncertainties.

The CSD slopes were estimated by the maximum likelihood method \citep[e.g.][]{Irwin+95,Loredo04}.
The differential surface density of asteroids with diameter $D$~km is represented by
\begin{equation}
\Sigma(D) = b~\Sigma(>\textrm{1~km})~D^{-b-1}.
\end{equation}
The likelihood function for $n$ objects is given by
\begin{equation}
 L = \exp \Bigl[ -\Omega \int{\eta(r(D))} \, \Sigma(D|b) \, \textrm{d}D \Bigr]
     \prod_{i=1}^{n} \sigma(D_i) \, \Sigma(D_i|b),
\end{equation}
where $\Omega$ is the survey area and $\sigma(D_i)$ is the uncertainty in diameter of object~$i$.
The function parameter of the detection efficiency $\eta$ is derived from that combined with the
selected data (see Section \ref{sec:analysis}).
$r(D)$ denotes the apparent $r$ magnitude converted from $D$ with $R$~=~2.9~AU (the mean of the
sample) and $p$~=~0.10 (weighted mean albedos in the whole main belt with $I$~$>$~15$\arcdeg$).

The likelihood analysis gives the slopes of $b$~=~1.25~$\pm$~0.03 ($1\sigma$) in
0.6~km $<$ $D$ $<$ 1.0~km and $b$~=~1.84~$\pm$~0.27 in 1.0~km $<$ $D$ $<$ 3.0~km.
The best-fit power-law CSDs are shown in figure \ref{fig9}.
We evaluated the fitting of the power laws using the Anderson-Darling statistic
\citep{Anderson&Darling52}, given by
\begin{equation}
 \textrm{AD} = \int_0^1 \frac{\bigl[ S(D) - P(D) \bigr]^2}{P(D) \bigl[ 1 - P(D) \bigr]} \,
               \textrm{d}P(D),
\end{equation}
where $P(D)$ the cumulative detection probability for an object larger than $D$ in diameter, and
$S(D)$ is the cumulative distribution function of the detected objects \citep{Bernstein+04}.
The goodness of fit is decided by the probability $Pr$(AD) of a random realization with AD-value
higher than the real data.
Low $Pr$(AD) (less than 0.05) implies a poor fit of the distribution.
Our calculation found $Pr$(AD)~=~0.52, indicating that the power-law fitting well represents the
observed CSD.

\cite{Terai&Itoh11} presented that $b$~=~1.79~$\pm$~0.05 for MBAs with $I$~$<$~15$\arcdeg$ and
$b$~=~1.62~$\pm$~0.07 for MBAs with $I$~$>$~15$\arcdeg$ in 0.7~km $<$ $D$ $<$ 2.0~km.
The slope for low-$I$ MBAs is much steeper than that of
\cite{Yoshida&Nakamura07}, $b$~=~1.29~$\pm$~0.02 in 0.6~km $<$ $D$ $<$ 1~km.
This discrepancy seems to be due to significant observational bias caused by the use of mixed
data taken in different ecliptic latitudes as well as solar phase angles.
In contrast, assuming that $b$ of high-$I$ MBAs is $\sim$0.1 smaller than that of low-$I$
MBAs as shown by \cite{Terai&Itoh11}, the result of this study, $b$~=~1.25 for high-$I$ MBAs, is
consistent with the CSD slope for low-$I$ sub-km MBAs given by \cite{Yoshida&Nakamura07}.
It shows a significant improvement in measurement accuracy of the size distribution
by an increase of the sample number, appropriate survey fields located around the opposition,
homogeneity of the survey region and data quality, and precise photometric calibration using the
background SDSS stars.

%%%%%%%%%%%%%%%%%%%%%%%%%%%%%%%%%%%%%%%%%%%%%%%%%%%%%%%%%%%%%%%%%%%%%%%%%%%%%%%%%%%%%%%%%%%%%%%%%%%
\subsection{Power-law Slope}\label{sec:powerlawslope}

Finally, we examined the difference in CSD slopes between low- and high-$I$ MBAs.
Previous asteroid surveys showed that the size distribution of MBAs exhibits a wavy pattern.
This structure is generated by the transition of impact strength between the strength- and
gravity-scaled regimes \citep{Davis+94,O'Brien&Greenberg03} as well as possibly by a small-size
cutoff due to the Poynting-Robertson drag and solar radiation pressure \citep{CampoBagatin+94}.
In addition, it has been indicated that the CSD slopes are different between S- and C-like
asteroids \citep{Ivezic+01,Yoshida&Nakamura07}.
In order to compare CSD slopes, the size range and number ratio of S- and C-like asteroids should
be conformed.

As a representative CSD slope of low-$I$ MBAs, we cited the results of \cite{Yoshida&Nakamura07}.
The colorimetric asteroid survey in the field within $\pm$3$\arcdeg$ from the ecliptic plane
detected a thousand of small MBAs, most of which have inclinations less than 10$\arcdeg$.
It presented the CSD slopes of $b$~=~1.29~$\pm$~0.02 in 0.3~km $<$ $D$ $<$ 1.0~km for S-like
asteroids and $b$~=~1.33~$\pm$~0.02 in 0.6~km $<$ $D$ $<$ 1.0~km for C-like asteroids.
\cite{Yoshida&Nakamura07} also showed the heliocentric distribution of the both classes.
In the outer region beyond $\sim$2.6~AU where most of the sample asteroids in our survey are
distributed, the fraction of S-like asteroids is $\sim$0.2 and the other is $\sim$0.8.

We conducted Monte Carlo simulations to estimate the CSD for low-inclination MBAs with given
fractions of the two groups.
Ten thousands of hypothetical S- and C-like asteroids are generated according to the abundance
ratio of 1:4.
Each asteroid is given diameter ranging from 0.6~km to 40~km following the differential size
distribution $\textrm{d}N/\textrm{d}D$~$\propto$~$D^{-b-1}$ with power-law indexes of
$-$2.29~$\pm$~0.02 for S-like asteroids and $-$2.33~$\pm$~0.02 for C-like asteroids. 
The result showed that the compound CSD obeys a power-law distribution with $b$~=~1.32~$\pm$~0.02
in 0.6~km $<$ $D$ $<$ 1.0~km.
It is significantly steeper than that obtained from our survey, $b$~=~1.25~$\pm$~0.03, with
the difference of $\Delta b$~=~0.07~$\pm$~0.04.
We confirmed that the high-inclination MBAs have a shallow CSD compared to the low-inclination
MBAs at least in sub-km size range.

On the other hand, it is difficult to compare the CSDs between \cite{Yoshida&Nakamura07} and this
study in the larger size range of $D$~$>$~1.0~km because of large uncertainty due to small number
of the samples and loss of some unmeasurable objects which are bright enough to reach saturation.
Instead, we analyzed the SDSS MOC including astrometric and photometric data for 471,569 moving
objects.
As in the case of our survey, most of them are unknown asteroids in orbit and albedo.
We estimate orbital elements and absolute magnitude of each SDSS MOC object using the same methods
and assumptions as this study (see section~\ref{sec:orbit} and \ref{sec:size}).

The SDSS MOC objects are classified according to the following definitions:
(i) MBAs are objects with $a$~=~2.1--3.3~AU.
(i\hspace{-.1em}i) S-like asteroids are objects with $a^{\ast}$~$>$~0, and C-like asteroids are
objects with $a^{\ast}$~$<$~0, where $a^{\ast}$ is the color index defined by
equation~(\ref{eq_aast}).
(i\hspace{-.1em}i\hspace{-.1em}i) Low-inclination asteroids are objects with $I$~$<$~15$\arcdeg$,
and high-inclination asteroids are objects with $I$~$>$~15$\arcdeg$.
Diameter of each MBA is estimated assuming $p$~=~0.22 for objects with $a^{\ast}$~$>$~0 as S-like
asteroids and $p$~=~0.07 for objects with $a^{\ast}$~$<$~0 as C-like asteroids.
The SDSS MOC seems to keep the complete detection up to $r$~=~21.2~mag corresponding to
$D$~$\leq$~2.0~km \citep{Parker+08}.
In 2.0~km $<$ $D$ $<$ 5.0~km, the CSDs of SDSS MOC MBAs have
$b$~=~2.65~$\pm$~0.03 for the low-inclination S-like,
$b$~=~2.17~$\pm$~0.04 for the high-inclination S-like,
$b$~=~2.24~$\pm$~0.02 for the low-inclination C-like, and
$b$~=~2.01~$\pm$~0.02 for the high-inclination C-like.
We confirmed that high-inclination MBAs have a shallower CSD in either class.

Then, model CSDs were generated from a mixture of the S-like and C-like MBAs with a number ratio
of 1:4 in each inclination population.
The estimation of CSD slopes in 2.0~km $<$ $D$ $<$ 5.0~km gives $b$~=~2.31~$\pm$~0.02 for the
low-inclination MBAs and $b$~=~2.04~$\pm$~0.02 for the high-inclination MBAs.
The difference of the slopes, $\Delta b$~=~$b$($I$~$<$~15$\arcdeg$)~$-$~$b$($I$~$>$~15$\arcdeg$),
is 0.27~$\pm$~0.03, much larger than the value indicated in our survey
($\Delta b$~=~0.07~$\pm$~0.04).
The discrepancy in $\Delta b$ between the two size ranges can be explained by the difference of
wavy pattern in CSDs.
We discuss the interpretations of this result in the following section.

%%%%%%%%%%%%%%%%%%%%%%%%%%%%%%%%%%%%%%%%%%%%%%%%%%%%%%%%%%%%%%%%%%%%%%%%%%%%%%%%%%%%%%%%%%%%%%%%%%%
\subsection{Impact Strength Law}\label{sec:impactstrengthlaw}

\cite{O'Brien&Greenberg03} presented an analytical model for steady-state size distributions
resulting from a collisional cascade with the following two essential facts.
First, the primary component of CSD slope represented by a single power-law index $b_p$ is given by
a simple expression of
\begin{equation}
b_p = \frac{5}{2+s/3},
 \label{eq_s2b}
\end{equation}
where $s$ is the power-law index of the $Q^\ast_D$ law, namely $Q^\ast_D$~$\propto$~$D^s$.
Second, the transition of the $Q^\ast_D$ law at a diameter of $D_t$~=~0.1--1.0~km induces the wavy
structure on the MBA size distribution.
In the gravity-scaled regime ($D$~$>$~$D_t$), the index $s_g$ is positive, i.e. $b_p$~$<$~2.5.
Conversely, in the strength-scaled regime ($D$~$<$~$D_t$), the index $s_s$ is negative, i.e.
$b_p$~$>$~2.5.
The inflection in the $Q^\ast_D$ law results wavelike oscillations about the CSD power law
with an index $b_p$ in the gravity-scaled regime.
The CSD shape in $D$~$>$~$D_t$ is determined by the primary slope $b_p$ as well as the phase and
amplitude of the wave pattern.
\cite{O'Brien&Greenberg05} found that the $Q^\ast_D$ law with $s_g$~$\approx$~1.40 and
$D_t$~$\approx$~0.2~km reproduces the observed MBA size distribution.

The diameter range of CSD measurements in this study is 0.7--5.0~km covering from the first bump
down from $D_t$.
\cite{Terai&Itoh11} confirmed no significant difference in the peak/valley positions of the
CSDs' wavy pattern between low- and high-inclination MBAs.
We suggest a simple model that the CSD shape varies only with $b_p$ and the wave amplitude between
the two populations.
The difference between a CSD slope in a local size range and the primary slope, $|b-b_p|$, is
assumed to shift in proportion to wave amplitude.
When the wave amplitude increases $k$~times, $b$ becomes $k(b-b_p)+b_p$.
This model allows to briefly express the relationships between $b_p$ and the estimated CSD slopes,
$b_{1,L}$, $b_{2,L}$, $b_{1,H}$, $b_{2,H}$, where the suffix 1 and 2 show the diameter ranges
of 0.6--1.0~km and 2.0--5.0~km, respectively, and the suffix L and H show the low- and
high-inclination populations, respectively

The MBAs' $Q^\ast_D$ law with $s_g$~=~1.40 \citep{O'Brien&Greenberg05} derives $b_p$~=~2.03 using
equation~(\ref{eq_s2b}).
It corresponds to the primary CSD slope of low-inclination MBAs.
The relational expressions of the CSD indexes in this model are given by
\begin{equation}
b_{i,H} - b_{p,H} = k(b_{i,L} - b_{p,L}), \quad i=1,2,
 \label{eq_model}
\end{equation}
where $b_{p,L}$ and $b_{p,H}$ are the primary CSD slopes for low- and high-inclination MBAs,
respectively ($b_{p,L}$~=~2.03).
Our analysis showed the local CSD slopes of
$b_{1,L}$~=~1.32~$\pm$~0.02, 
$b_{2,L}$~=~2.31~$\pm$~0.02, 
$b_{1,H}$~=~1.25~$\pm$~0.03, and
$b_{2,H}$~=~2.04~$\pm$~0.02.
Those give the solution of equation~(\ref{eq_model}) that $b_{p,H}$~=~1.82~$\pm$~0.07 and
$k$~=~0.80~$\pm$~0.04.
This $b_{p,H}$ value is converted into $s_g$~=~2.2~$\pm$~0.3 when equation~(\ref{eq_s2b}) is
applied.
However, the collisional evolution of high-inclination MBAs is dominated by collisions with
low-inclination asteroids though equation~(\ref{eq_model}) is based on collisional equilibrium
in a self-contained system \citep{O'Brien&Greenberg03}.
Numerical simulations for the collisional evolution are required to derive $s_g$ from $b_{p,H}$.
But anyway, the result in $b_{p,H}$~$<$~$b_{p,L}$ indicates a steep $Q^\ast_D$ law in the
gravity-scaled regime of high-inclination MBAs.
It leads to the conclusion that hypervelocity collisions on large bodies are relatively less
disruptive.
This may suggest that the inelasticity parameter determining the fraction of impact energy
partitioned into fragment kinetic energy, generally denoted by $f_{\rm KE}$
\citep{CampoBagatin+01,O'Brien&Greenberg05}, decreases with collision velocity around several
km~sec$^{-1}$.

Our results imply that the collisional evolution and the resulting size distribution of an
asteroid population suffered hypervelocity collisions are not the same as those of ecliptic MBAs
even if the compositions and internal structures are similar to each other.
In the inner region of planetesimal disk after formation of giant planets and gas dissipation,
small bodies are dynamically excited and collide with each another at high velocities.
\cite{Bottke+05b} showed that in the primordial main belt zone during the dynamical excitation
phase caused by planetary embryos and Jupiter, collisions between remnant asteroids occur at
6--8~km~sec$^{-1}$ and collisions between remnant asteroids and depleted asteroids reach more than
10~km~sec$^{-1}$. 
The velocity-dependent $Q^\ast_D$ law should be introduced for investigating the ancient size
distribution and its evolution of asteroids at the final stage of planet formation processes.
It allows to more precisely estimate the impact rate and size distribution of meteorites colliding
with the Earth and moon in the early solar system.

Besides high-inclination MBAs, near-Earth asteroids (NEAs) collide with each other at greater than
10~km~sec$^{-1}$ \citep{Bottke+94}.
Jovian Trojans with high inclinations ($I$~$\gtrsim$~20$\arcdeg$) also have mean collision
velocities of $\sim$6~km~sec$^{-1}$ or higher \citep{Marzari+96}.
In addition, high-velocity collisional processes were confirmed in several planetesimal disks
such as HD172555 \citep{Lisse+09} and Epsilon Eridani systems \citep{Thebault+02}.

The relationship between $Q^\ast_D$ law and collision velocity is required to be determined
by combination of further studies with observations, numerical simulations, and laboratory
experiments.
It provides insight into the collisional evolution of small-body populations located in the regions
close to the host star or distant from the ecliptic plane, as well as in the systems around a
massive star or containing giant planets.

%%%%%%%%%%%%%%%%%%%%%%%%%%%%%%%%%%%%%%%%%%%%%%%%%%%%%%%%%%%%%%%%%%%%%%%%%%%%%%%%%%%%%%%%%%%%%%%%%%%
%%%%%%%%%%%%%%%%%%%%%%%%%%%%%%%%%%%%%%%%%%%%%%%%%%%%%%%%%%%%%%%%%%%%%%%%%%%%%%%%%%%%%%%%%%%%%%%%%%%
\section{CONCLUSIONS}\label{sec:conclusions}

Our survey detected 441 asteroids in 13.6~deg$^2$ at high ecliptic latitudes with a 50\% limiting
magnitude of $r$~=~24.4~mag.
We obtained an unbiased sample consisting of 221 MBA candidates with inclination of
$I$~$\gtrsim$~14$\arcdeg$ and absolute magnitude of $H_r$~$\leq$~19.4~mag. 
Although orbits and diameters of each asteroid cannot be determined and instead are estimated
with the assumption of a circular orbit and given albedo depending on radial regions, the sample
yields a sufficiently quality CSD to measure its slope in sub-km size.

The CSD for high-inclination MBAs shows a roll-over at $D$~$\sim$~1.0~km which has also been
indicated by \cite{Yoshida&Nakamura07} for low-inclination MBAs.
The maximum likelihood analysis provided the best-fit power laws with $b$~=~1.25~$\pm$~0.03
in 0.6~km $<$ $D$ $<$ 1.0~km and $b$~=~1.84~$\pm$~0.27 in 1.0~km $<$ $D$ $<$ 3.0~km.
Most of the MBA candidates are located beyond 2.6 AU where \cite{Yoshida&Nakamura07} showed the
abundance ratio of S- and C-like asteroids is 1:4 for low-inclination sub-km MBAs.
The compound CSD with the number ratio of 1:4 has $b$=1.32~$\pm$~0.02 in 0.6~km $<$ $D$ $<$ 1.0~km,
indicating that high-inclination MBAs have a shallower CSD in sub-km size.

We furthermore examined the CSDs in 2.0~km $<$ $D$ $<$ 5.0~km using the SDSS MOC database.
The slopes are $b$~=~2.31~$\pm$~0.02 for low-inclination MBAs and $b$~=~2.04~$\pm$~0.02 for
high-inclination asteroids.
Although high-inclination MBAs have shallower CSDs in both of the size ranges, the slope difference
is larger in the larger size.
This inconsistency is explained to be due to the difference of the wavy pattern on CSDs between
low- and high-inclination populations.
Assuming a simple model that the both populations have the same positions of `bump' on
the CSDs at a few-km diameter, high-inclination MBAs have a primary CSD slope of
$b_p$=1.82~$\pm$~0.07 over the diameter range from 0.6~km to 5.0~km.
It is definitely shallower than that of low-inclination MBAs, indicating that hypervelocity
collisions raises the relative strength of large bodies against catastrophic impacts.

\acknowledgments

We thank Melissa McGrath for providing valuable comments.
This study is based on data collected at Subaru Telescope.
T. Terai was supported by the Grant-in-Aid from Japan Society for the Promotion of Science
(20-4879).

{\it Facilities:} \facility{Subaru Telescope (NAOJ)}.

%%%%% References %%%%%

\clearpage
%%%%% Figure 1 %%%%%

\clearpage
%%%%% Figure 2 %%%%%

\clearpage
%%%%% Figure 3 %%%%%

\clearpage
%%%%% Figure 4 %%%%%

\clearpage
%%%%% Figure 5 %%%%%

\clearpage
%%%%% Figure 6 %%%%%

\clearpage
%%%%% Figure 7 %%%%%

\clearpage
%%%%% Figure 8 %%%%%

\clearpage
%%%%% Figure 9 %%%%%


\begin{thebibliography}{}

\bibitem[Anderson \& Darling(1952)]{Anderson&Darling52} Anderson T. W., \& Darling, D. A.\ 1952,
 Annals of Mathematical Statistics, 23, 193
\bibitem[Benz \& Asphaug(1999)]{Benz&Asphaug+99} Benz, W., \& Asphaug, E.\ 1999, \icarus, 142, 5
\bibitem[Bernstein et al.(2004)]{Bernstein+04} Bernstein, G.~M., Trilling, D.~E., Allen, R.~L.,
 et al.\ 2004, \aj, 128, 1364
\bibitem[Bertin \& Arnouts(1996)]{Bertin&Arnouts96} Bertin, E., \& Arnouts, S.\ 1996, \aaps, 117,
 393
\bibitem[Bottke et al.(1994)]{Bottke+94} Bottke, W.~F., Jr., Nolan, M.~C., Greenberg, R.,
 \& Kolvoord, R.~A.\ 1994, Hazards Due to Comets and Asteroids, 337 
\bibitem[Bottke et al.(2005a)]{Bottke+05a} Bottke, W.~F., Durda, D.~D., Nesvorn{\'y}, D., et al.
 \ 2005, \icarus, 175, 111 
\bibitem[Bottke et al.(2005b)]{Bottke+05b} Bottke, W.~F., Durda, D.~D., Nesvorn{\'y}, D., et al.
 \ 2005, \icarus, 179, 63 
\bibitem[Bowell et al.(1989)]{Bowell+89} Bowell, E., Hapke, B., Domingue, D., et al.\ 1989,
 Asteroids II, 524
\bibitem[Bowell et al.(1994)]{Bowell+94} Bowell, E., Muinonen, K., \& Wasserman, L.~H.\ 1994,
 Asteroids, Comets, Meteors 1993, 160, 477
\bibitem[Bus \& Binzel(2002)]{Bus&Binzel02} Bus, S.~J., \& Binzel, R.~P.\ 2002, \icarus, 158, 146 
\bibitem[Campo Bagatin et al.(1994)]{CampoBagatin+94} Campo Bagatin, A., Cellino, A., Davis, D.~R.,
 Farinella, P., \& Paolicchi, P.\ 1994, \planss, 42, 1079
\bibitem[Campo Bagatin et al.(2001)]{CampoBagatin+01} Campo Bagatin, A., Petit, J.-M.,
 \& Farinella, P.\ 2001, \icarus, 149, 198 
\bibitem[Davis et al.(1994)]{Davis+94} Davis, D.~R., Ryan, E.~V., \& Farinella, P.\ 1994, \planss,
 42, 599 
\bibitem[Davis et al.(2002)]{Davis+02} Davis, D.~R., Durda, D.~D., Marzari, F., Campo Bagatin, A., 
 \& Gil-Hutton, R.\ 2002, Asteroids III, 545 
\bibitem[de El{\'{\i}}a \& Brunini(2007)]{deElia&Brunini07} de El{\'{\i}}a, G.~C., \& Brunini, A.
 \ 2007, \aap, 466, 1159 
\bibitem[Durda et al.(1998)]{Durda+98} Durda, D.~D., Greenberg, R., \& Jedicke, R.\ 1998,
 \icarus, 135, 431 
\bibitem[Farinella \& Davis(1992)]{Farinella&Davis+92} Farinella, P., \& Davis, D.~R.\ 1992,
 \icarus, 97, 111 
\bibitem[Fukugita et al.(1996)]{Fukugita+96} Fukugita, M., Ichikawa, T., Gunn, J. E., Doi, M.,
 Shimasaku, K., \& Schneider, D. P.\ 1996, \aj, 111, 1748
\bibitem[Fukugita et al.(2011)]{Fukugita+11} Fukugita, M., Yasuda, N., Doi, M., Gunn, J.~E.,
 \& York, D.~G.\ 2011, \aj, 141, 47
\bibitem[Gil-Hutton(2006)]{Gil-Hutton06} Gil-Hutton, R.\ 2006, \icarus, 183, 93 
\bibitem[Gladman et al.(1998)]{Gladman+98} Gladman, B., Kavelaars, J.~J., Nicholson, P.~D., Loredo,
 T.~J., \& Burns, J.~A.\ 1998, \aj, 116, 2042 
\bibitem[Gladman et al.(2009)]{Gladman+09} Gladman, B.~J., Davis, D.~R., Neese, C., et al.\ 2009,
 \icarus, 202, 104 
\bibitem[Housen \& Holsapple(1990)]{Housen&Holsapple90} Housen, K.~R., \& Holsapple, K.~A.\ 1990,
 \icarus, 84, 226 
\bibitem[Ivezi{\'c} et al.(2001)]{Ivezic+01} Ivezi{\'c}, {\v Z}., Tabachnik, S., Rafikov, R.,
 et al.\ 2001, \aj, 122, 2749 
\bibitem[Ivezi{\'c} et al.(2002)]{Ivezic+02} Ivezi{\'c}, {\v Z}., Juri\'c, M., Lupton, R.~H.,
 Tabachnik, S., \& Quinn, T.\ 2002, \procspie, 4836, 98
\bibitem[Irwin et al.(1995)]{Irwin+95} Irwin, M., Tremaine, S., \& Zytkow, A.~N.\ 1995, \aj, 110,
 3082
\bibitem[Jedicke(1996)]{Jedicke96} Jedicke, R.\ 1996, \aj, 111, 970 
\bibitem[Jedicke \& Metcalfe(1998)]{Jedicke&Metcalfe+98} Jedicke, R., \& Metcalfe, T.~S.\ 1998,
 \icarus, 131, 245
\bibitem[Jutzi et al.(2010)]{Jutzi+10} Jutzi, M., Michel, P., Benz, W., \& Richardson, D.~C.\ 2010,
 \icarus, 207, 54 
\bibitem[Kadono et al.(2010)]{Kadono+10} Kadono, T., Sakaiya, T., Hironaka, Y., et al.\ 2010,
 Journal of Geophysical Research (Planets), 115, 4003 
\bibitem[Lisse et al.(2009)]{Lisse+09} Lisse, C.~M., Chen, C.~H., Wyatt, M.~C., et al.\ 2009,
 \apj, 701, 2019 
\bibitem[Loredo(2004)]{Loredo04} Loredo, T.~J.\ 2004, American Institute of Physics Conference
 Series, 735, 195 
\bibitem[Marzari et al.(1996)]{Marzari+96} Marzari, F., Scholl, H., \& Farinella, P.\ 1996,
 \icarus, 119, 192 
\bibitem[Masiero et al.(2011)]{Masiero+11} Masiero, J.~R., Mainzer, A.~K., Grav, T., et al.\ 2011,
 \apj, 741, 68 
\bibitem[Miyazaki et al.(2002)]{Miyazaki+02} Miyazaki, S., Komiyama, Y., Sekiguchi, M.,
 et al.\ 2002, \pasj, 54, 833
\bibitem[Nakamura \& Yoshida(2002)]{Nakamura&Yoshida02} Nakamura, T., \& Yoshida, F.\ 2002,
 \pasj, 54, 1079
\bibitem[O'Brien \& Greenberg(2003)]{O'Brien&Greenberg03} O'Brien, D.~P., \& Greenberg, R.\ 2003,
 \icarus, 164, 334 
\bibitem[O'Brien \& Greenberg(2005)]{O'Brien&Greenberg05} O'Brien, D.~P., \& Greenberg, R.\ 2005,
 \icarus, 178, 179 
\bibitem[Ouchi et al.(2004)]{Ouchi+04} Ouchi, M., Shimasaku, K., Okamura, S., et al.\ 2004, \apj,
 611, 660
\bibitem[Parker et al.(2008)]{Parker+08} Parker, A., Ivezi{\'c}, {\v Z}., Juri{\'c}, M., et al.
 \ 2008, \icarus, 198, 138
\bibitem[Petit et al.(2001)]{Petit+01} Petit, J.-M., Morbidelli, A., \& Chambers, J.\ 2001,
 \icarus, 153, 338 
\bibitem[Takasawa et al.(2011)]{Takasawa+11} Takasawa, S., Nakamura, A.~M., Kadono, T., et al.
 \ 2011, \apjl, 733, L39 
\bibitem[Tedesco et al.(2002)]{Tedesco+02} Tedesco, E.~F., Noah, P.~V., Noah, M., \& Price, S.~D.
 \ 2002, \aj, 123, 1056 
\bibitem[Terai et al.(2007)]{Terai+07} Terai, T., Itoh, Y., \& Mukai, T.\ 2007, \pasj, 59, 1175
\bibitem[Terai \& Itoh(2011)]{Terai&Itoh11} Terai, T., \& Itoh, Y.\ 2011, \pasj, 63, 335
\bibitem[Th{\'e}bault et al.(2002)]{Thebault+02} Th{\'e}bault, P., Marzari, F., \& Scholl, H.
 \ 2002, \aap, 384, 594 
\bibitem[Usui et al.(2011)]{Usui+11} Usui, F., Kuroda, D., M{\"u}ller, T.~G., et al.\ 2011,
 \pasj, 63, 1117
\bibitem[van Houten et al.(1970)]{vanHouten+70} van Houten, C.~J., van Houten-Groeneveld, I.,
 Herget, P., \& Gehrels, T.\ 1970, \aaps, 2, 339 
\bibitem[Vedder(1998)]{Vedder+98} Vedder, J.~D.\ 1998, \icarus, 131, 283 
\bibitem[Yoshida et al.(2003)]{Yoshida+03} Yoshida, F., Nakamura, T., Watanabe, J.-I., et al.
 \ 2003, \pasj, 55, 701 
\bibitem[Yoshida \& Nakamura(2007)]{Yoshida&Nakamura07} Yoshida, F., \& Nakamura, T.\ 2007,
 \planss, 55, 1113


\end{thebibliography}
\end{document}